\documentclass{article}
\usepackage{fullpage}
\usepackage[authoryear, round]{natbib}
\usepackage{amsmath}
\usepackage{enumitem}
\include{comment}
\usepackage{url}
\usepackage{booktabs}
\usepackage{mathtools}
\mathtoolsset{showonlyrefs}
\usepackage{float}

\begin{document}

\title{Unified Host and Network Data Set}
\author{Melissa J. M. Turcotte$^{*}$, Alexander D. Kent$^{*}$ and Curtis Hash$^{\dagger}$\\
  $^{*}$Los Alamos National Laboratory,
Los Alamos, NM, 87545, U.S.A.\\ 
mturcotte@lanl.gov\\
$^{\dagger}$Ernst \& Young}
\date{}
\maketitle

\begin{abstract}
  The lack of data sets derived from operational enterprise networks
  continues to be a critical deficiency in the cyber security research
  community. Unfortunately, releasing viable data sets to the larger
  community is challenging for a number of reasons, primarily the
  difficulty of balancing security and privacy concerns against the
  fidelity and utility of the data. This chapter discusses the importance
  of cyber security research data sets and introduces a large data set
  derived from the operational network environment at Los Alamos National
  Laboratory. The hope is that this data set and associated discussion
  will act as a catalyst for both new research in cyber security as well
  as motivation for other organizations to release similar data sets to
  the community. 
\end{abstract}

\section{Introduction}

The lack of diverse and useful data sets for cyber security research continues
to play a profound and limiting role within the relevant research communities
and their resulting published research. Organizations are reticent to release
data for security and privacy reasons. In addition, the data sets that are
released are encumbered in a variety of ways, from being stripped of so much
information that they no longer provide rich research and analytical
opportunities, to being so constrained by access restrictions that key details
are lacking and independent validation is difficult. In many cases,
organizations do not collect relevant data in sufficient volumes or with high
enough fidelity to provide cyber research value. Unfortunately, there is
generally little motivation for organizations to overcome these obstacles.

In an attempt to help stimulate a larger research effort focused on operational
cyber data as well as to motivate other organizations to release useful data
sets, Los Alamos National Laboratory (LANL) has released two data sets for
public use \citep{kent2016, kent2014}. A third, entitled the ``Unified Host and
Network Data Set," is introduced in this chapter.

The Unified Host and Network Data Set is a subset of network flow and computer
events collected from the LANL enterprise network over the course of
approximately 90 days.\footnote{The network flow data are only 89 days due to
missing data on the first day.} The host (computer) event logs originated from
the majority of LANL's computers that run the Microsoft Windows operating
system. The network flow data originated from many of the internal core routers
within the LANL enterprise network and are derived from router netflow records.
The two data sets include many of the same computers but are not fully
inclusive; the network data set includes many non-Windows computers and other
network devices.

Identifying values within the data sets have been de-identified (anonymized) to
protect the security of LANL's operational IT environment and the privacy of
individual users. The de-identified values match across both the host and
network data allowing the two data elements to be used together for analysis
and research. In some cases, the values were not de-identified, including
well-known network ports, system-level usernames (not associated to people) and
core enterprise hosts. In addition, a small set of hosts, users and processes
were combined where they represented well-known, redundant entities. This
consolidation was done for both normalization and security purposes.

In order to transform the data into a format that is useful for researchers who
are not domain experts, a significant effort was made to normalize the data
while minimizing the artifacts that such normalization might introduce.

\subsection{Related public data sets}
A number of public, cyber security relevant data sets currently exist
\cite{glasserbridging, darpadata, impactdata, malwaredata, ma2009identifying,
canadadata}.  Some of these represent data collected from operational
environments, while others capture specific, pseudo real-world events (for
example, cyber security training exercises). Many data sets are synthetic and
created using models intended to represent specific phenomenon of relevance;
for example, the Carnegie Melon Software Engineering Institute provides several
insider threat data sets that are entirely synthetic \cite{glasserbridging}. In
addition, many of the data sets commonly seen within the research community are
egregiously dated. The DARPA cyber security data sets \cite{darpadata}
published in the 1990s are still regularly used, even though the systems,
networks and attacks they represent have almost no relevance to modern
computing environments. 

Another issue is that many of the available data sets have restrictive access
and constraints on how they may be used. For example, the U.S. Department of
Homeland Security provides the Information Marketplace for Policy and Analysis
of Cyber-risk and Trust (IMPACT) \cite{impactdata}, which is intended to
facilitate information sharing. However, the use of any of the data hosted by
IMPACT requires registration and vetting prior to access. In addition, data
owners may (and often do) place limitations on how and where the data may be
used.

Finally, many of the existing data sets are not adequately characterized for
potential researchers. It is important that researchers have a thorough
understanding of the context, normalization processes, idiosyncracies and other
aspects of the data. Ideally, researchers should have sufficiently detailed
information to avoid making false assumptions and to reproduce similar data.
The need for such detailed discussion around published data sets is a primary
purpose of this chapter.

The remainder of this chapter is organized as follows: a description of the
Network Flow Data is given in Section \ref{sec:network} followed by the Windows
Host Log Data in Section \ref{sec:wls}. Finally, a discussion of potential
research directions is given in Section \ref{sec:research}.

\section{Network Flow Data}\label{sec:network}
The network flow data set included in this release is comprised of records
describing communication events between devices connected to the LANL
enterprise network. Each \textit{flow} is an aggregate summary of a (possibly)
bi-directional network communication between two network devices. The data are
derived from Cisco NetFlow Version 9 \cite{rfc3954} flow records exported by
the core routers. As such, the records lack the payload-level data upon which
most commercial intrusion detection systems are based. However, research has
shown that flow-based techniques have a number of advantages and are successful
at detecting a variety of malicious network behaviors
\cite{sperotto2010overview}. Furthermore, these techniques tend to be more
robust against the vagaries of attackers, because they are not searching for
specific signatures (e.g., byte patterns) and they are encryption-agnostic.
Finally, in comparison to full-packet data, collection, analysis and archival
storage of flow data at enterprise scales is straightforward and requires
minimal infrastructure.

\subsection{Collection \& Transformation}\label{sec:collectandtransform}
As mentioned previously, the raw data consisted of NetFlow V9 records that were
exported from the core network routers to a centralized collection server.
While V9 records can contain many different fields, only the following are
considered: \textit{StartTime}, \textit{EndTime}, \textit{SrcIP},
\textit{DstIP}, \textit{Protocol}, \textit{SrcPort}, \textit{DstPort},
\textit{Packets} and \textit{Bytes}. The specifics of the hardware and flow
export protocol are largely irrelevant, as these fields are common to all
network flow formats of which the authors are aware.

This data can be quite challenging to model without a thorough understanding of
its various idiosyncrasies. The following paragraphs discuss two of the most
relevant issues with respect to modelling. For a comprehensive overview of
these issues, among others, readers can refer to \cite{hofstede2014flow}.

Firstly, note that these flow records are uni-directional (\textit{uniflows}):
each record describes a stream of packets sent from one network device
(\textit{SrcIP}) to another (\textit{DstIP}). Hence, an established TCP
connection --- bi-directional by definition --- between two network devices,
\textit{A} and \textit{B}, results in two flow records: one from \textit{A} to
\textit{B} and another from \textit{B} to \textit{A}. It follows that there is
no relationship between the direction of a flow and the initiator of a
bi-directional connection (i.e., it is not known whether \textit{A} or
\textit{B} connected first). This is the case for most netflow implementations
as bi-directional flow (\textit{biflow}) protocols such as \cite{rfc5103} have
yet to gain widespread adoption. Clearly, this presents a challenge for
detection of attack behaviors, such as lateral movement, where directionality
is of primary concern.

Secondly, significant duplication can occur due to flows encountering multiple
netflow sensors in transit to their destination. Routers can be configured to
track flows on ingress and egress, and, in more complex network topologies, a
single flow can traverse multiple routers. More recently, the introduction of
netflow-enabled switches and dedicated netflow appliances has exacerbated the
issue. Ultimately, a single flow can result in many distinct flow records. To
add further complexity, the flow records are not necessarily \textit{exact}
duplicates and their arrival times can vary considerably; these inconsistencies
occur for many reasons, the particulars of which are too complex to discuss in
this context.

\begin{table}{
    \centering
  \begin{tabular*}{1\textwidth}{@{\extracolsep{\fill}} lp{0.65\textwidth}}
    \toprule[\heavyrulewidth]
    {\bf Field Name} & {\bf Description}\\
    \midrule[\heavyrulewidth]
        {\it Time} & The start time of the event in epoch time format.\\
        {\it Duration} & The duration of the event in seconds.\\
        {\it SrcDevice} & The device that likely initiated the event.\\
        {\it DstDevice} & The receiving device.\\
        {\it Protocol} & The protocol number.\\
        {\it SrcPort} & The port used by the \textit{SrcDevice}.\\
        {\it DstPort} & The port used by the \textit{DstDevice}.\\
        {\it SrcPackets} & The number of packets the \textit{SrcDevice} sent during the event.\\
        {\it DstPackets} & The number of packets the \textit{DstDevice} sent during the event.\\
        {\it SrcBytes} & The number of bytes the \textit{SrcDevice} sent during the event.\\
        {\it DstBytes} & The number of bytes the \textit{DstDevice} sent during the event.\\
        \bottomrule[\heavyrulewidth]
     
  \end{tabular*}}
 \caption{Bi-directional flow data}
  \label{tab:netflow}
\end{table}

In order to simplify the data for modelling, a transformation process known as
\textit{biflowing} or \textit{stitching} was employed. This is a process
intended to aggregate duplicates and marry the opposing uniflows of
bi-directional connections into a single, \textit{directed} biflow record
(Table \ref{tab:netflow}). Many approaches to this problem can be found in the
literature \cite{minarik2009improving, nguyen2017flow, barbosa2014anomaly,
berthier2010nfsight}, all of them imperfect.  A straightforward approach was
used that relies on simple port heuristics to decide direction. These
heuristics are based on the assumption that \textit{SrcPort}s are generally
\textit{ephemeral} (i.e., they are selected from a pre-defined, high range by
the operating system), while \textit{DstPort}s tend to have lower numbers that
correspond to established, shared network services and will therefore be
observed more frequently than ephemeral ports. The heuristics are given below in
order of precedence.

\begin{itemize}
	\item Destination ports are less than 1024 and source ports are not.
	\item The top 90 most frequently observed ports are destination ports.
	\item The smaller of the two ports is the destination port.
\end{itemize}

Each uniflow was transformed into a biflow by renaming the \textit{Packets} and
\textit{Bytes} fields to \textit{SrcPackets} and \textit{SrcBytes}
respectively. \textit{DstPackets} and \textit{DstBytes} fields were added with
initial values of zero. Next, the port heuristics were considered and, if any
were violated or ambiguous, the \textit{Src} and \textit{Dst} attributes were
swapped, effectively reversing the direction. Finally, the \textit{5-tuple} was
extracted from each record and used as the key in a lookup table.

% could mention that we failed to limit this to tcp. it sort of works for some
% udp protocols like DNS. makes no sense for icmp.

\begin{center}
	\textit{SrcIP}, \textit{DstIP}, \textit{SrcPort}, \textit{DstPort}, \textit{Protocol}
\end{center}

If a match was found, the flows were aggregated by keeping the minimum
\textit{StartTime}, maximum \textit{EndTime} and summing the other attributes.
If no match was found, the flow was simply added to the table. This process was
performed in a streaming fashion on all of the records in the order in which
they were received by the collector. Flows were periodically evicted from the
lookup table after 30 minutes of inactivity (i.e., failing to match with any
incoming flows). Flows that remained active for long periods of time were
reported approximately every 3 hours, but were \textit{not} evicted from the
table until inactive.

While biflowing the data mitigates the problems posed by duplicates and
ambiguous directionality, it does not address another significant obstacle: the
lack of stable identifiers upon which to build models. In some cases, IP
addresses are transient (e.g., DHCP, VPN). In other cases, devices have
multiple IP addresses (e.g., multihoming) or one IP address is shared by
multiple devices (e.g., load-balancing, NAT). Whatever the case may be,
modelling the behavior of IP addresses on a typical network is clearly error
prone. Instead, one should endeavor to map IP addresses to more stable
identifiers such as Media Access Control (MAC) addresses or fully-qualified
domain names (FQDN), interchangeably referred to as hostnames throughout the
rest of the chapter. As with directionality, there is no perfect solution to
this problem. The most appropriate identifier will depend greatly on the
configuration of the target network, as well as the availability of auxiliary
data sources from which a mapping can be constructed. An ideal solution will
likely involve some combination of supplementary network data (e.g., DNS logs,
DNS zone transfers, DHCP logs, VPN logs, NAC logs), business rules and
considerable trial and error.

% a little acronym soup there ... do we need to spell all of those out?

For this data release, a combination of Domain Name Service (DNS) and Dynamic
Host Configuration Protocol (DHCP) logs was used to construct a mapping of IP
addresses to FQDNs over time. The IP addresses in each biflow were then
replaced with their corresponding FQDNs at the time of the flow. Where a given
IP address and timestamp mapped to multiple FQDNs, business rules were
incorporated to give preference to the least-ephemeral name. IP addresses that
failed to map to any FQDN were left as-is. The resulting mix of names and IP
addresses correspond to the \textit{SrcDevice} and \textit{DstDevice} fields in
the final data.

Finally, the data were de-identified by mapping \textit{SrcDevice},
\textit{DstDevice}, \textit{SrcPort} and \textit{DstPort} to random
identifiers. In the event that the IP-to-FQDN mapping failed, the random
identifier was prepended with ``IP." Well-known ports were not de-identified.
Records with protocol numbers other than 6 (TCP), 17 (UDP) and (1) ICMP were
removed entirely. The output from this process is provided in CSV format, one
record per line, with fields in the order shown in Table \ref{tab:netflow}.

\subsection{Data Quality}
\begin{figure}[t!]
\centerline{
\includegraphics[width = 1\textwidth]{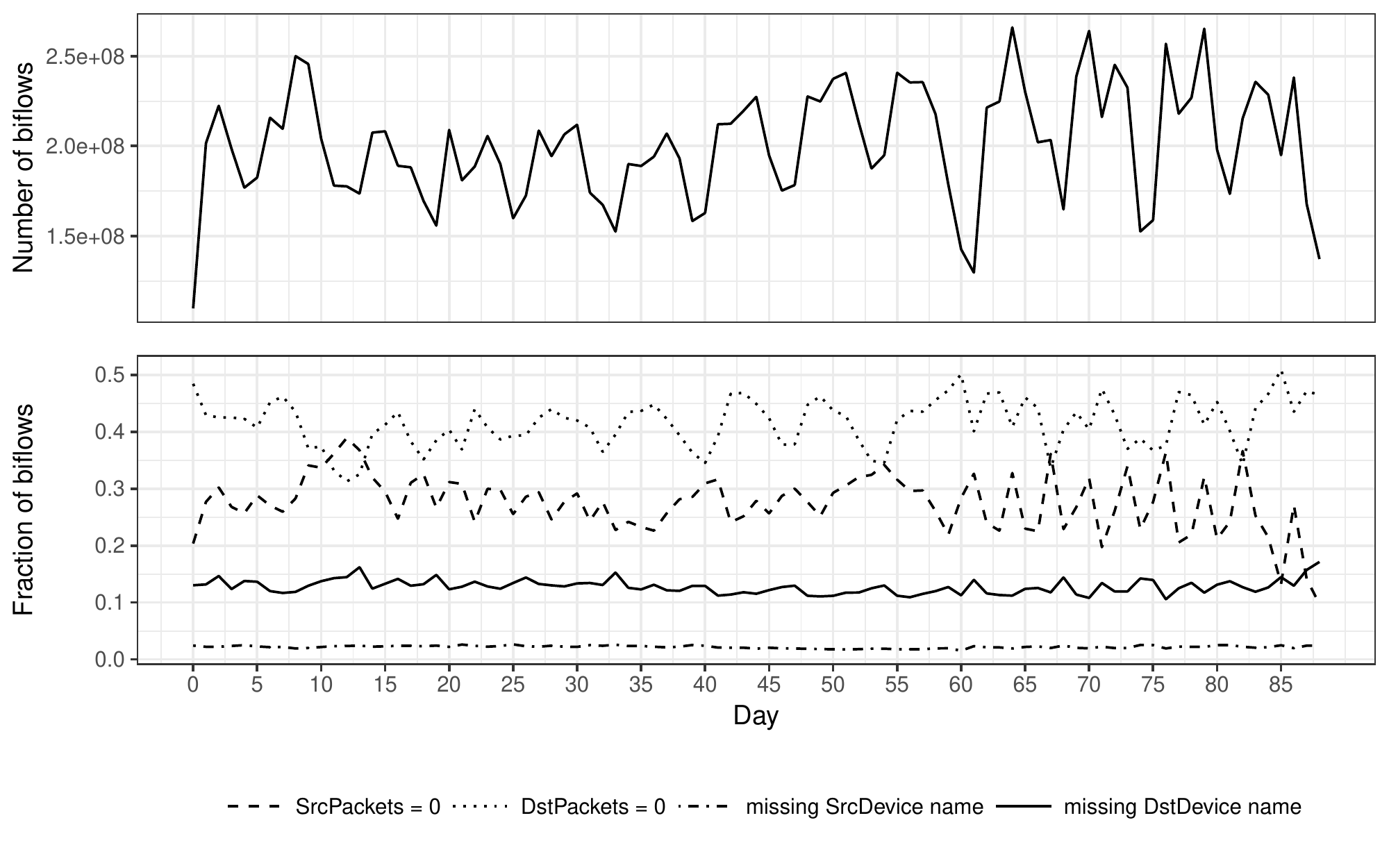}
}
\caption{\textbf{Top}: Daily count of biflows by end time. \textbf{Bottom}:
	Fraction of biflows where \textit{SrcPackets} = 0, \textit{DstPackets}
	= 0, \textit{SrcDevice} FQDN-mapping failed and \textit{DstDevice}
	FQDN-mapping failed.}
\label{fig:dailynetflow}
\end{figure}

Several figures have been provided in order to assess the quality of the
network flow data set. The top plot in Figure \ref{fig:dailynetflow}, which
shows the number of biflows over time, demonstrates the periodicity that one
would expect for data whose volume is driven by the comings and goings of
employees during a typical 5-day workweek.

The bottom plot of Figure \ref{fig:dailynetflow} is intended to measure the
success rate of the biflowing and IP-to-FQDN mapping processes. TCP biflows
where either \textit{SrcPackets} or \textit{DstPackets} is zero suggests a
failure to find matching uniflows for both directions of the exchange. 57\% of
TCP and approximately 70\% of all biflows fall within this category. This can
largely be attributed to LANL's netflow sensor infrastructure, which has been
specifically configured to export only one direction on many routes. In
addition, some devices --- namely vulnerability scanners and the like ---
attempt to connect to all possible IP addresses within a range; this results in
a significant number of uniflows for which no response is possible. Likely for
the same reason, IP-to-FQDN mapping failed for significantly more
\textit{DstDevice}s than \textit{SrcDevice}s.

\begin{figure}[h!]
\centerline{
\includegraphics[width = 1\textwidth]{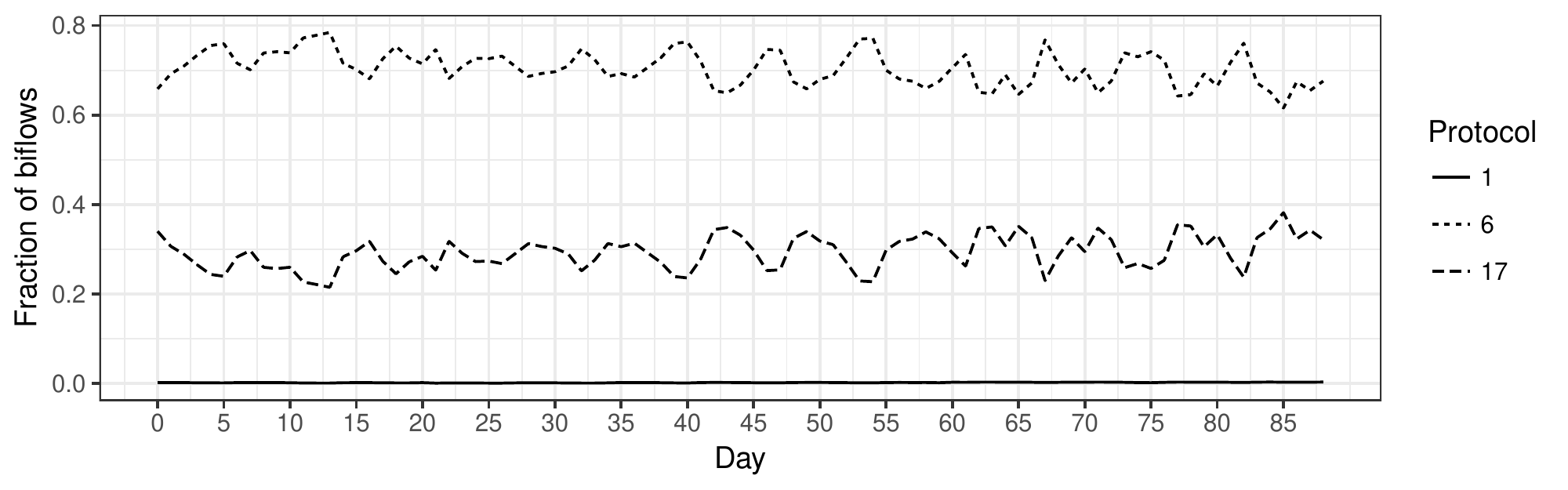}
}
\caption{Daily proportions of each \textit{Protocol}.}
\label{fig:dailyprotocol}
\end{figure}

\begin{figure}[t!]
\centerline{
\includegraphics[width = 1\textwidth]{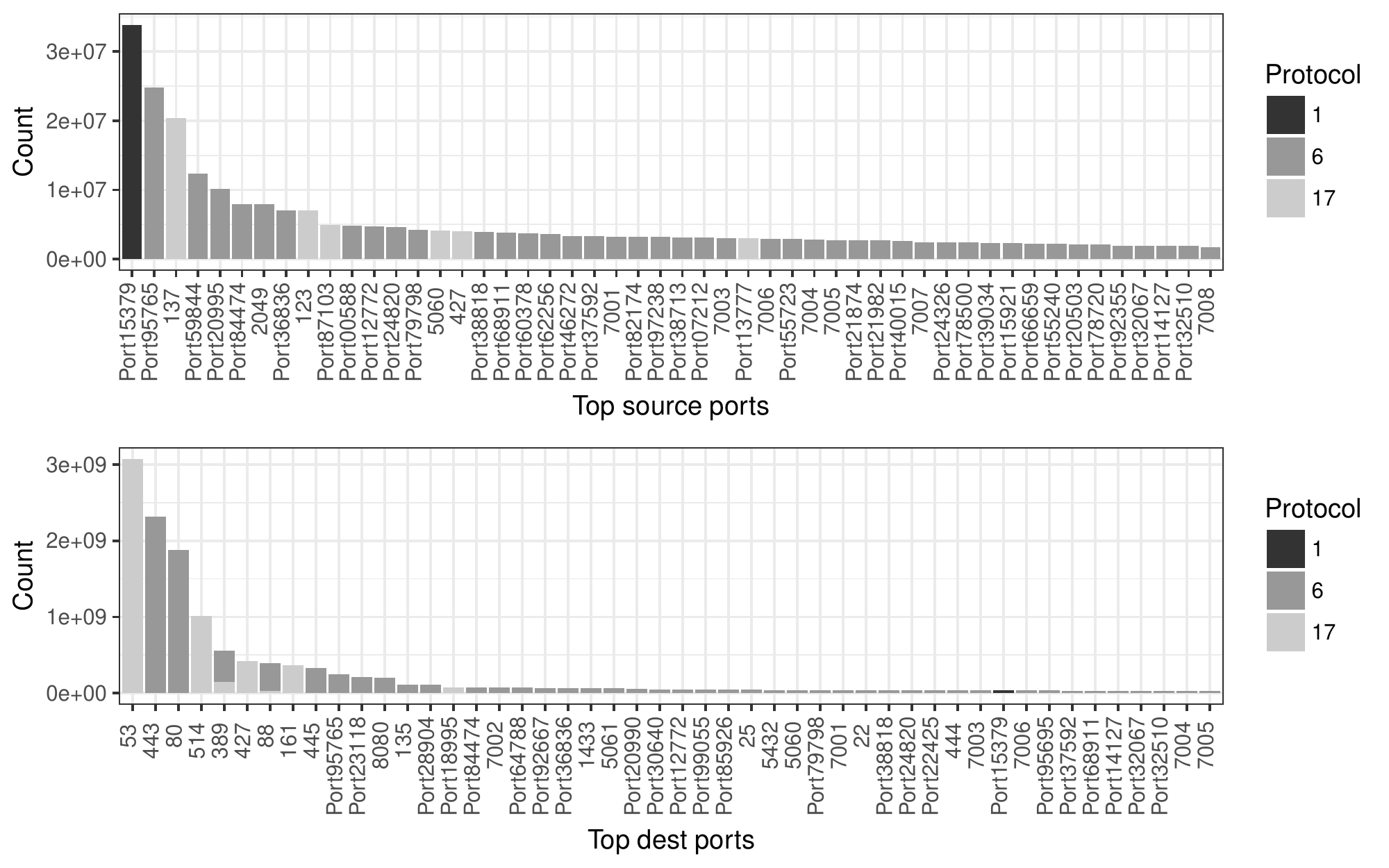}
}
\caption{Histogram of the top 50 \textit{SrcPort}s and \textit{DstPort}s.}
\label{fig:topports}
\end{figure}

Figure \ref{fig:dailyprotocol} shows the daily proportion of biflows
corresponding to each \textit{Protocol}. Figure \ref{fig:topports} contains two
histograms of the top \textit{SrcPort}s and \textit{DstPort}s respectively.
Note the non-uniformity in the \textit{SrcPort} histogram; this illustrates
either a consistent failure of the biflowing process to choose the appropriate
direction or the presence of protocols that use non-ephemeral source ports. For
example, the Network Time Protocol (NTP) uses port 123 for both the source and
destination ports per the specification.

\begin{figure}[H]
\centerline{
\includegraphics[width = 1\textwidth]{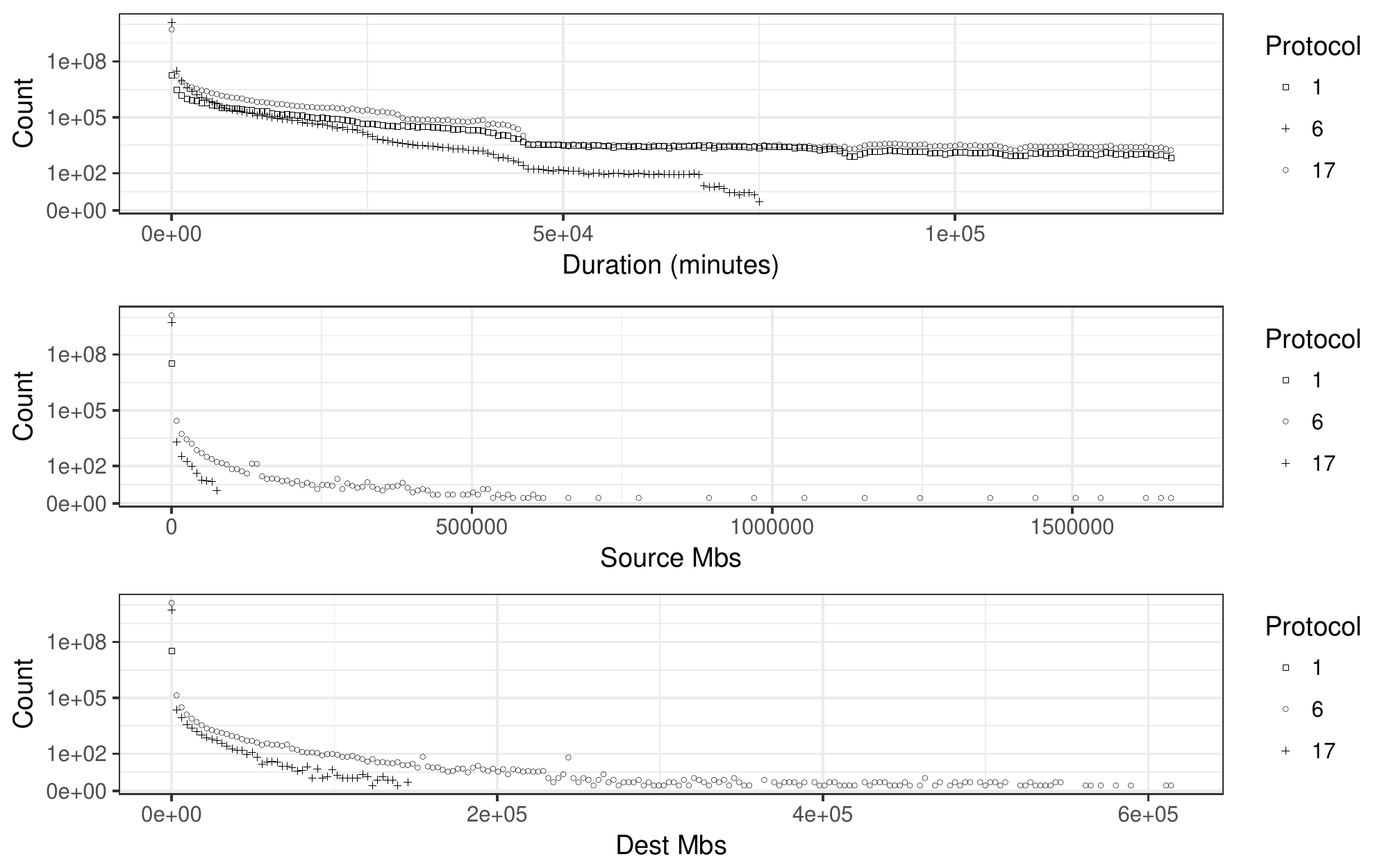}
}
\caption{Distribution of \textit{Duration}, \textit{SrcBytes} and \textit{DstBytes}.}
\label{fig:bytesduration}
\end{figure}

\begin{figure}[ht!]
\centerline{
\includegraphics[width = 1\textwidth]{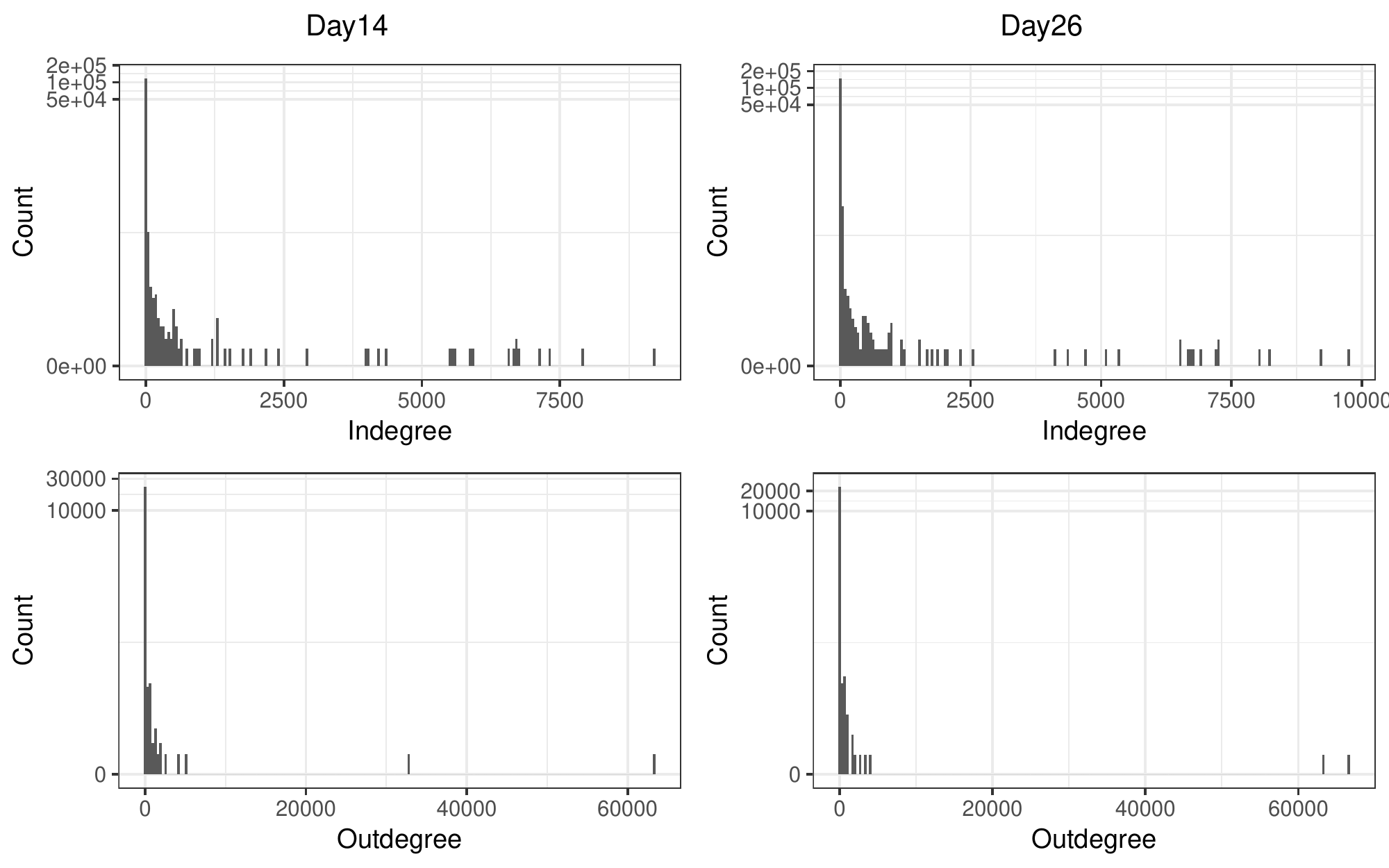}
}
\caption{In-degree and out-degree distribution for two randomly-selected days.}
\label{fig:degreedist}
\end{figure}

Figure \ref{fig:bytesduration} shows the distribution of \textit{Duration},
\textit{SrcBytes} and \textit{DstBytes} per \textit{Protocol}. Of particular
interest is the presence of many long-lived UDP and ICMP biflows in the data.
This indicates frequent, persistent UDP and ICMP traffic sharing the same
\textit{5-tuple} and is an unfortunate side-effect of not limiting the biflow
transformation to TCP uniflows. Finally, Figure \ref{fig:degreedist} shows
exemplar in-degree and out-degree distributions for two randomly-selected days.


\section{Windows Host Log Data}\label{sec:wls}

As remote attackers and malicious insiders increasingly use encryption,
network-only detection mechanisms are becoming less effective, particularly
those that require the inspection of payload data within the network traffic.
As a result, cyber defenders now rely heavily on endpoint agents and host event
logs to detect and investigate incidents. Host event logs capture nuanced
details for a wide range of activities; however, given the vast number of
logged events and their specificity to an individual host, human analysts
struggle to discover the few useful log entries amid the huge number of
innocuous entries. Statistical analytics for host event data are in their
infancy. Advanced analytical capabilities on this host data, including computer
and user profiling, which move beyond signature-based methods, will increase
network awareness and detection of advanced cyber threats.

The host event data set is a subset of host event logs collected from all
computers running the Microsoft Windows operating system on LANL's enterprise
network. The host logs were collected with Windows Logging Service (WLS), which
is a Windows service that forwards event logs, along with administrator-defined
contextual data to a set of collection servers \cite{wls}. The released data
are in JSON format in order to preserve the structure of the original events,
unlike the two previously released data sets based on this log source
\cite{kent2016, kent2014}. The events from the host logs included in the data
set are all related to authentication and process activity on each machine. 

\begin{figure}[h!]
\centerline{
\includegraphics[width = 1\textwidth]{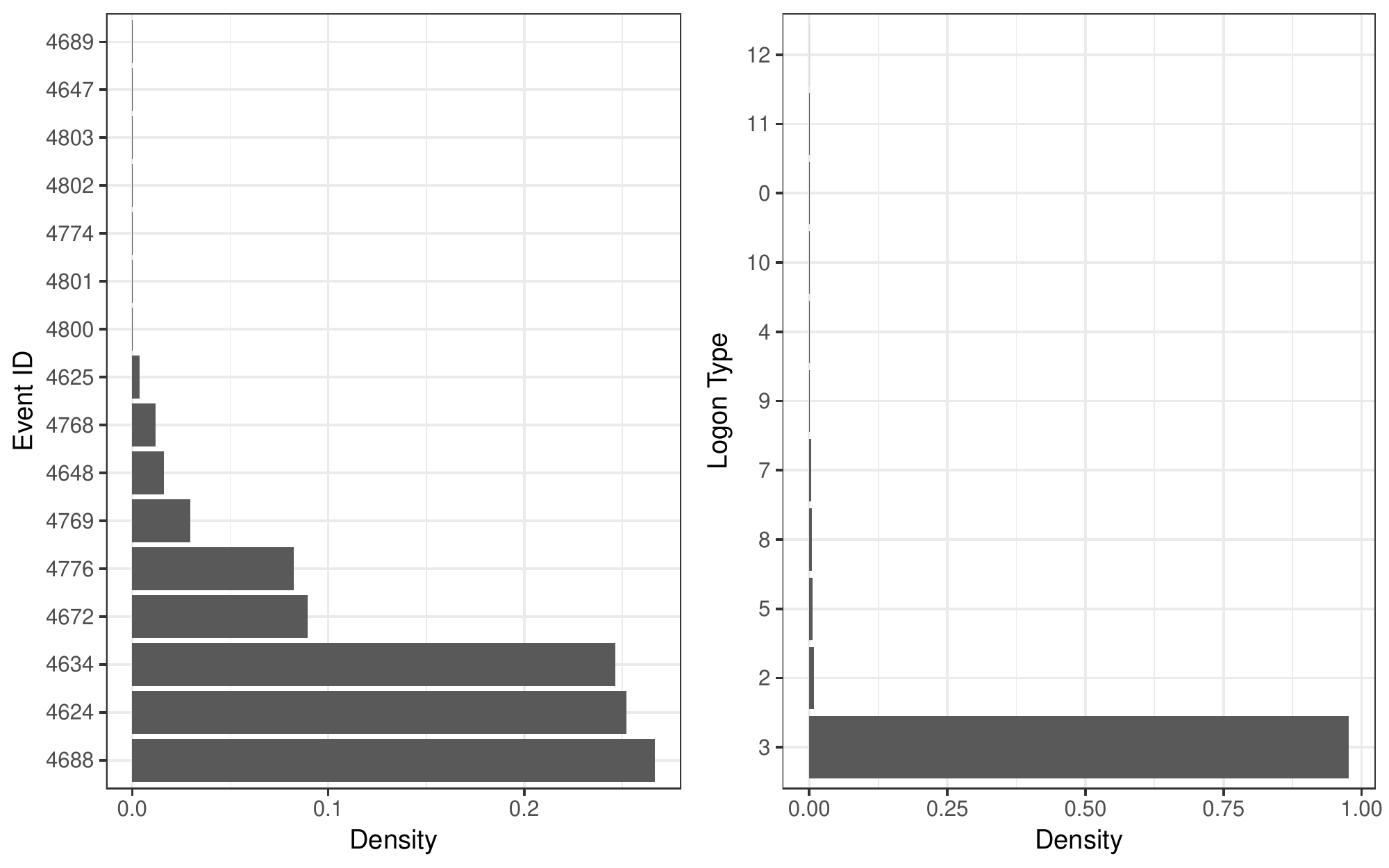}
}
\caption{Histogram of the Event IDs and Logon Types.}
\label{fig:wlshist}
\end{figure}

Table \ref{tab:eventid} contains the subset of \textit{EventID}s included from
the event logs in the released data set and a brief description of each; see
\cite{ultimatewindows} for a more detailed description. Figure
\ref{fig:wlshist} shows the percentage of \textit{EventID}s contained in the
logs, as well as the \textit{LogonType}s for \textit{EventID}s 4624, 4625 and
4634.

\begin{table}[h!]
  \centering
  {\begin{tabular*}{1\textwidth}{@{\extracolsep{\fill}} lll}
    \toprule[\heavyrulewidth]
    Event ID & Description\\
    \midrule[\heavyrulewidth]
    \multicolumn{3}{c}{Authentication events}\\
    \midrule
    4768 & \multicolumn{2}{l}{Kerberos authentication ticket was requested (TGT)}\\
    4769 & \multicolumn{2}{l}{Kerberos service ticket was requested (TGS)}\\
    4770 & \multicolumn{2}{l}{Kerberos service ticket was renewed}\\
    4774 & \multicolumn{2}{l}{An account was mapped for logon}\\
    4776 & \multicolumn{2}{l}{The domain controller attempted to validate credentials}\\
    4624 & \multicolumn{2}{l}{An account was successfully logged on, see Logon Types}\\
    4625 & \multicolumn{2}{l}{An account failed to logon, see Logon Types}\\
    4634 & \multicolumn{2}{l}{An account was logged off, see Logon types}\\
    4647 & \multicolumn{2}{l}{User initiated logoff}\\
    4648 & \multicolumn{2}{l}{A logon was attempted using explicit credentials}\\
    4672 & \multicolumn{2}{l}{Special privileges assigned to a new logon}\\
    4800 & \multicolumn{2}{l}{The workstation was locked}\\
    4801 & \multicolumn{2}{l}{The workstation was unlocked}\\
    4802 & \multicolumn{2}{l}{The screensaver was invoked}\\
    4803 & \multicolumn{2}{l}{The screensaver was dismissed}\\
    \midrule
    \multicolumn{3}{c}{Process events}\\
    \midrule
    4688 & \multicolumn{2}{l}{Process start}\\
    4689 & \multicolumn{2}{l}{Process end}\\
    \midrule
    \multicolumn{3}{c}{System events}\\
    \midrule
    4608 & \multicolumn{2}{l}{Windows is starting up}\\
    4609 & \multicolumn{2}{l}{Windows is shutting down}\\
    1100 & \multicolumn{2}{l}{Event logging service has shut down}\\
    \mbox{}&\multicolumn{2}{l}{(often recorded instead of EventID 4609)}\\
    \toprule[\heavyrulewidth]
    \multicolumn{3}{c}{Logon Types (EventIDs: 4624, 4625 and 4634)}\\
    \midrule
    2 - Interactive & 5 - Service & 9\phantom{0} - NewCredentials\\
    3 - Network & 7 - Unlock & 10 - RemoteInteractive\\
    4 - Batch & 8 - NetworkClearText & 11 - CachedInteractive\\
    12 - CachedRemoteInteractive & \multicolumn{2}{l}{0 - Used only by the system account}\\
    \bottomrule[\heavyrulewidth]
  \end{tabular*}}
  \caption{Host log Event IDs}
  \label{tab:eventid}
\end{table}

Each record in the data set will have some of the event attributes listed in
Appendix \ref{app:fields} and the table in Appendix \ref{app:attributes}
specifies which Event IDs have each attribute. Note that not all events with a
given \textit{EventID} share the same set of attributes. If an expected
attribute was missing from the original host log record, then the attribute was
not included in the corresponding record in the de-identified data set.

\begin{figure}[t!]
\centerline{
\includegraphics[width = 1\textwidth]{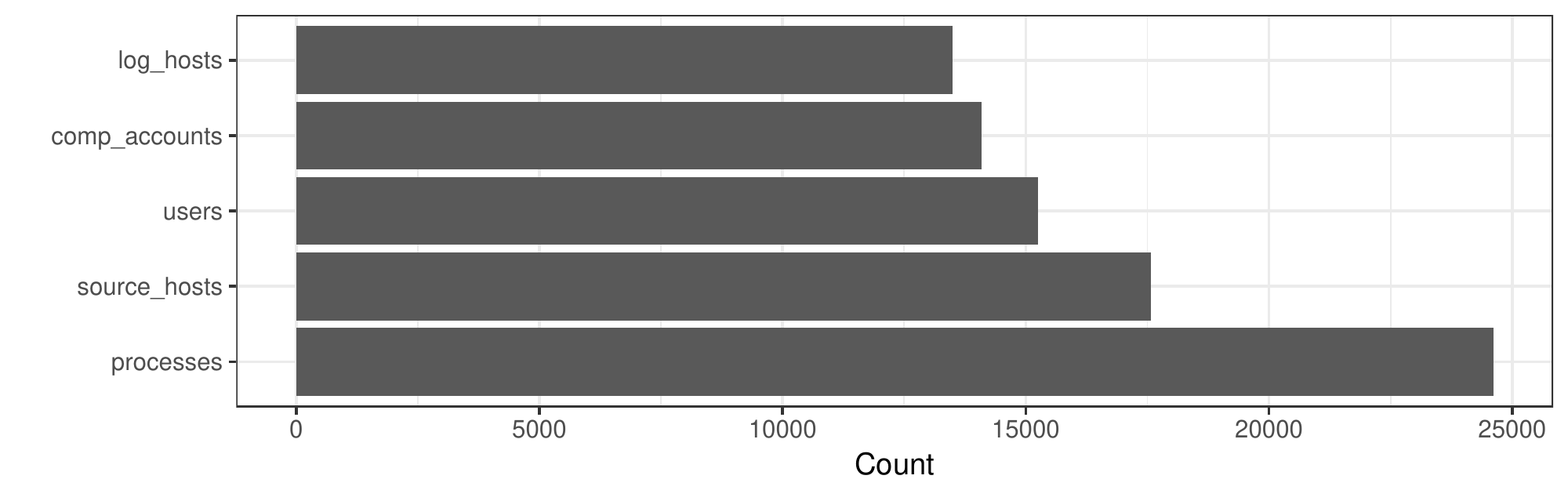}
}
\caption{Histogram of unique processes, usernames, log hosts
  ({\it LogHost}), source hosts ({\it Source}) and
  computer accounts for the whole time period.}
\label{fig:wlshistfields}
\end{figure}

\begin{figure}[h!]
\centerline{
\includegraphics[width = 1\textwidth]{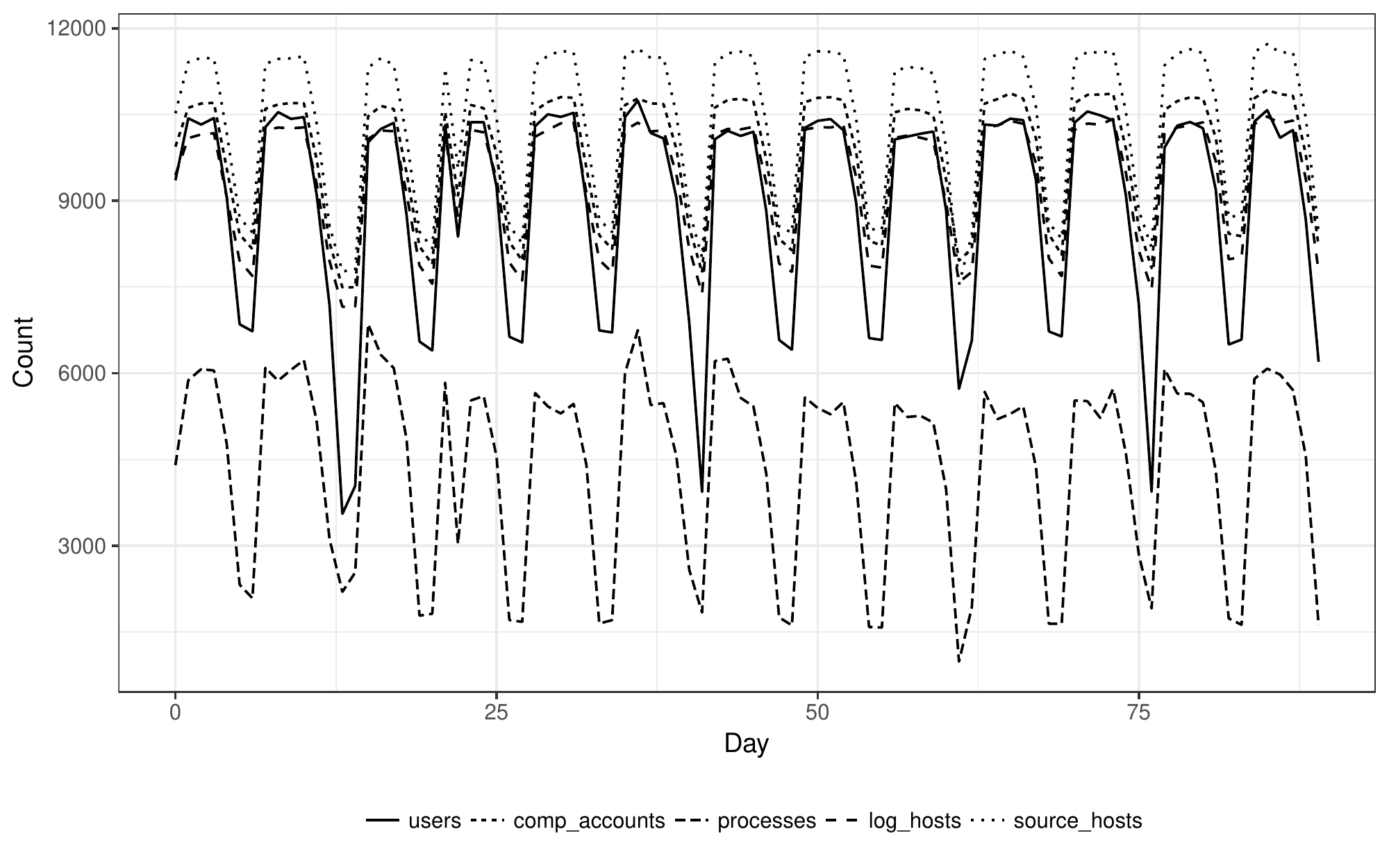}
}
\caption{Daily count of the fields in Figure \ref{fig:wlshistfields}.}
\label{fig:wlscountsday}
\end{figure}

All records will contain the attributes {\it EventID}, {\it LogHost} and {\it
Time}. {\it LogHost} indicates the network host where the record was logged.
For directed authentication events, this attribute will always correspond to
the computer to which the user is authenticating, and the source computer will
be given by {\it Source}. For the user associated with the record, if the {\it
UserName} ends in \$ then it will correspond to the \textit{computer account}
for the specified computer. These computer accounts are host-specific accounts
within the Microsoft Active Directory domain that allow the computer to
authenticate as a unique entity within the network. Figure
\ref{fig:wlshistfields} shows the count of unique processes, log hosts ({\it LogHost}), source hosts ({\it Source}), computer accounts ({\it UserName}
ending in \$) and users ({\it UserName} not ending in \$) for the 90 day
period. Figure \ref{fig:wlscountsday} shows the count for the same attributes
on a per-day basis. Note that the set of source hosts includes devices running
non-Windows operating systems, hence there are more source hosts than log
hosts.

Requests to the Kerberos Ticket Granting Service (TGS) (\textit{EventID} 4769)
correspond to a user requesting Kerberos authentication credentials from the
Active Directory domain to a service or account name on a network computer.
Hence, the {\it LogHost} attribute should always be an Active Directory
machine and the service or account name the user is requesting access to will
be given by {\it ServiceName}. The \textit{ServiceName} often corresponds to a
computer account on the target computer. Because this event only grants a
credential, a subsequent network logon event (\textit{EventID} 4624 -
\textit{LogonType} 3) to the computer indicated by \textit{ServiceName} is
common. This differs from the previous data release \cite{kent2016}, in which
TGS events were assumed to be directed authentication events from the user's
machine to the computer indicated by \textit{ServiceName}, ignoring the
Kerberos intermediary. 

When de-identifying the process events, only the base process name was
de-identified and the extension was left as is. Further, the parent process
names ({\it ParentProcessName}) do not have file extensions unlike the child
process names ({\it ProcessName}); this is a direct artifact of how the process
information is logged within WLS. The missing extension can be obtained by
using the \textit{ParentProcessID} to identify the parent process start event.

Finally, many events include the {\it DomainName} attribute that indicates what
Active Directory domain the event is associated with. The domain, combined with
the \textit{UserName}, should be considered a unique account identity. For
example, user \emph{u1} with domain \emph{d1} is not necessarily user \emph{u1}
in domain \emph{d2}. In addition, the domain may actually be a hostname,
indicating the event does not involve a user or account associated with an
Active Directory domain, but is instead a local account. Again, these accounts
should be considered unique to the host indicated within the
\textit{DomainName} attribute. For example, the Administrator account on host
\emph{c1} likely does not have a relationship to the Administrator account on
\emph{c2} or the Administrator account in domain \emph{d1}. The LANL data sets
have a single primary domain, with a number of much smaller, secondary domains
and most computers have a small set of local accounts. 

\subsection{Data parsing considerations}
While host logs can be an extremely valuable data resource for cyber security
research, the formatting and content of the logs can vary drastically between
enterprises depending upon the audit policy and technologies used to collect
and forward the logs to a centralized server. Hence, parsing the data and
extracting the relevant attributes is an important first step in analyzing
these data; see also \cite{kent2016}.

Even though WLS provides more content and normalization around the raw Windows
logs, some challenges were still faced to provide the de-identified data.

Firstly, the semantics of attribute names are not necessarily the same for
different \textit{EventID}s and the attribute names themselves may differ
according to what tool is being used to collect and forward the logs. For
example, with WLS the {\it UserName} for \textit{EventID} $4774$ is {\it
MappedName}, for EventID $4778$ and $4779$ it is {\it AccountName} and for most
other events it is {\it TargetUserName}. When parsing the data, these names
were all standardized to {\it UserName}.

As with the network flow data, an extremely important task is mapping IP
addresses to FQDNs. Further, unlike netflow, each record may contain both IP
addresses and hostnames. The machine where the event is recorded ({\it
LogHost} for the de-identified data) is provided as a hostname, whereas the
\textit{Source} computer for network logons is often given as an IP address.

Finally, both usernames and process names were standardized. In some records,
usernames appear with the domain name or additional characters. These
discrepancies were removed from the released data in order to ensure all
usernames were in canonical form. In addition, some usernames, such as
``Anonymous'', ``Local Service'' and ``Network Service'', do not map to a
computer or user account. For some analyses, one may want to remove these
events. In the de-identified data these commonly-seen usernames were not
anonymized. For the process names, dates, version numbers, operating systems
and hexadecimal strings were removed where possible so that processes run on
different operating systems or with different versions would map to the same
process name. For example, {\it flashplayerplugin\_20\_0\_0\_286.exe} would be
mapped to {\it flashplayerplugin\_VERSION.exe}. 

\section{Research directions}\label{sec:research}
Anomaly detection for the defensive cyber domain is a major yet evolving
research area, with much work still to be done in characterizing and finding
anomalies within complex cyber data sets. Finding viable attack indicators and
per computer, user and computer-to-computer models that enable anomaly
detection and fingerprinting are all interesting and important research
opportunities.

Although research on anomaly detection for cyber defense spans more than two
decades, operational tools are still almost exclusively rule- or
signature-based. Two reasons that statistical methods have not been more widely
adopted in practice are a high false positive rate and un-interpretable alerts.
Analysts are inundated with a large number of alerts and triaging them takes
significant time and resources; this results in low tolerance for false alarms
and alerts that provide no contextual information to guide investigation.
Signature-based systems can be finely tuned to reduce false positives as they
rely on very specific peculiarities that have been previously identified and
documented as indicative of a cyber attack. Further, they are interpretable as
they refer to specific patterns within the data, such as weird domains, network
protocols or process names.

However, despite their inherent challenges, anomaly detection methods have the
advantage of being able to detect new variants of cyber attacks and are able to
keep pace with the rapidly changing cyber attack landscape by dynamically
learning patterns for normal behavior and detecting deviations.  Further, with
the increasing level of encrypted network traffic, the importance of this
research and the use of these methods can not be understated. Research into
ways to reduce false positives and providing interpretable anomalies will have
significant impact in furthering the use of anomaly detection systems. In fact,
providing interpretable anomalies can help overcome the false positive issue as
interpretability leads to quickly identifying alerts that are false positives
in the same way it would enable understanding true positives. Research
approaches to tackle these problems could include combining different data sets
and signals, borrowing strength across entities that are similar by
incorporating peer-based behavior, community detection approaches and ways to
provide meaningful context surrounding alerts to human analysts.

\begin{figure}[t!]
\centerline{
\includegraphics[width = 1\textwidth]{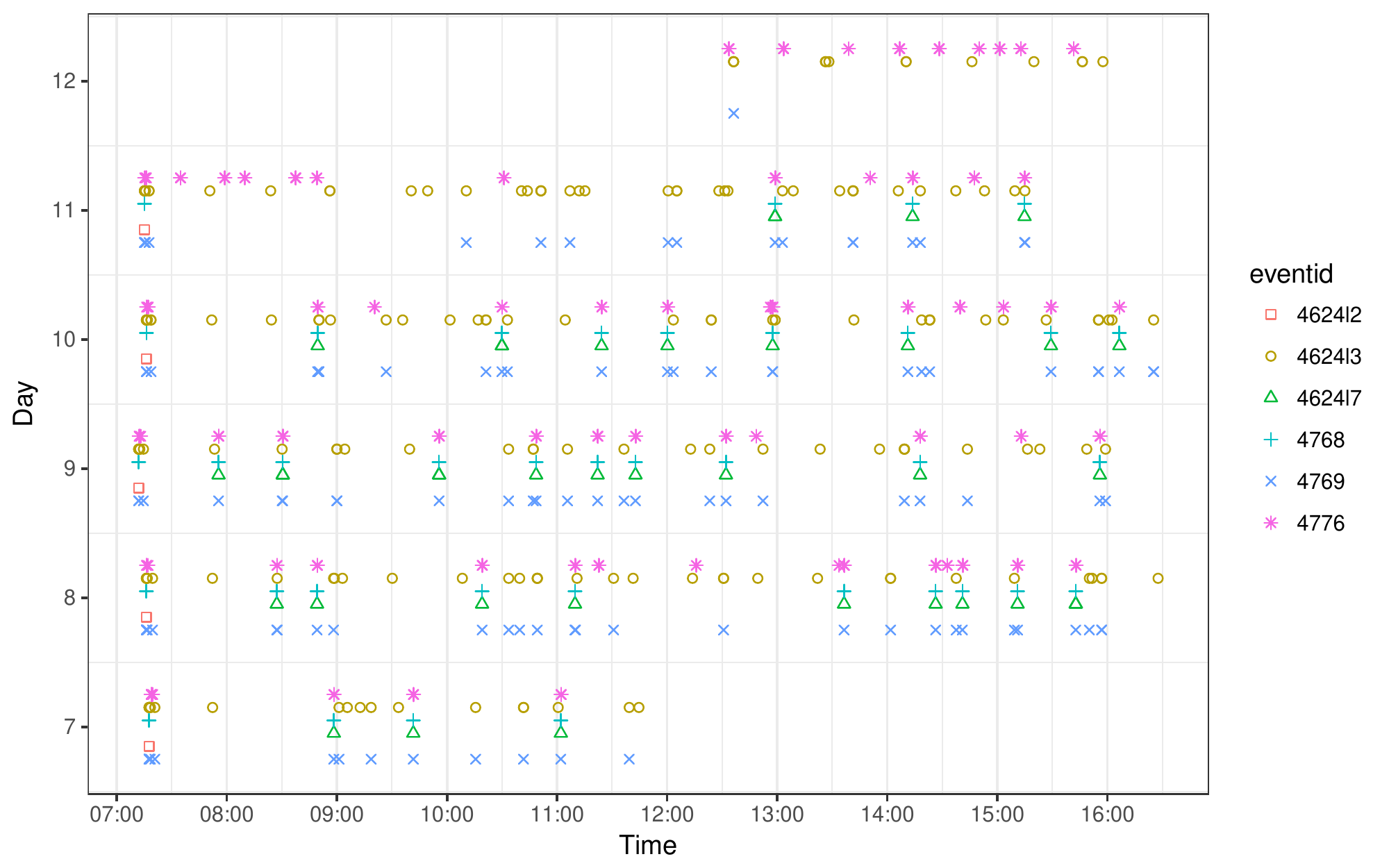}
}
\caption{Event times for User205265. $4624$l$2$ corresponds to \textit{EventID} 4624 - \textit{LogonType} 2.}
\label{fig:usereventtimes}
\end{figure}

When using the host log data set for research, some notable characteristics of
these data that need to be considered, especially if looking at the events as a
time series, is periodicity and significant correlations between arrivals of
different event types. This can be seen clearly in Figure
\ref{fig:usereventtimes}, which shows the event times for various
\textit{EventID}s for User205265. Periodicity in the data is often an
artifact of the computer regularly renewing credentials. This explains why
\textit{EventID} 4624 - \textit{LogonType} 3 (network logon) constitutes such a
significant portion of the events as seen in Figure \ref{fig:wlshist}. For a
given entity, extrapolating higher-level, interpretable actions from the
sequence of low-level events would improve modelling efforts, understanding of
these data and would itself be very useful for security analysts. See
\cite{heardpolling} and \cite{mpw} for relevant research in this
area.

Another area for research with the host logs is exploring the records related
to process starts and stops in detail, in particular looking at process trees.
To date, little has been done in this area. Computer systems operate
hierarchically; an initial root process starts many other processes, which in
turn start and run descendants. A process tree is the dynamic structure that
results. In theory, any process can be traced, through its ancestors, to the
root process. Unusual or atypical events in process trees could indicate
potential cyber security anomalies.

Moving beyond anomaly detection, there are other important research directions
for which these data could prove useful. For example, preliminary work has been
done using similar data to model network segmentation and associated risk
\cite{pope17}. Using the data to build new, potential network topologies in
order to reduce risk and improve security posture are viable directions.
Another potential research problem is to quantify and understand data loss
within cyber data sets. The collection and normalization processes in place for
these data can result in information loss and understanding this data loss is
an open problem both in general and specific to each element of the data. As
most of the data elements represent people and their actions on computers,
research on organizational and social behavior is also viable using these data.

\section{Conclusion}
Operational cyber security data sets are paramount to ensuring valuable and
productive research continues to improve the state of cyber defense. The
network flow and host log event data discussed in this chapter are intended to
enable such research as well as to provide an example for other potential data
set providers. In particular, while there is a considerable amount of relevant
work on network data, relatively little attention has been given to host log
data in the literature. Host log data are becoming increasingly relevant as
endpoint security tools gain popularity within the cyber security ecosystem.
It is important that researchers embrace both the opportunity and challenge
that they present. Finally, even less consideration has been given to
meaningful analyses that combine these and other data sets. This paradigm shift
towards a holistic approach to cyber security defense is critical to advancing
the state of the art.

\section{Acknowledgment}
This work has been authored by an employee of Los Alamos National Security,
LLC, operator of the Los Alamos National Laboratory under Contract No.
DE-AC52-06NA25396 with the U.S. Department of Energy. The United States
Government retains and the publisher, by accepting this work for publication,
acknowledges that the United States Government retains a nonexclusive, paid-up,
irrevocable, world-wide license to publish or reproduce this work, or allow
others to do so for United States Government purposes.
\appendix
\section{Host log fields}\label{app:fields}
    \begin{itemize}[leftmargin=*]
\item {\it Time:} The epoch time of the event in seconds.
\item {\it EventID:} Four digit integer corresponding to the event id of the record.
\item {\it LogHost:} The hostname of the computer that the event was recorded on. In the case of directed authentication events, the {\it LogHost} will correspond to the computer that the authentication event is terminating at (destination computer).
\item {\it LogonType:} Integer corresponding to the type of logon, see Table \ref{tab:eventid}.
\item {\it LogonTypeDescription:} Description of the {\it LogonType}, see Table \ref{tab:eventid}.
\item {\it UserName:} The user account initiating the event. If the user ends in $\$$, then it corresponds to a computer account for the specified computer.
\item {\it DomainName:} Domain name of {\it UserName}. 
\item {\it LogonID:} A semi-unique (unique between current sessions and {\it LogHost}) number that identifies the logon session just initiated. Any events logged subsequently during this logon session should report the same Logon ID through to the logoff event.
\item {\it SubjectUserName:} For authentication mapping events, the user account specified by this field is mapping to the user account in {\it UserName}.
\item {\it SubjectDomainName:} Domain name of {\it SubjectUserName}.
\item {\it SubjectLogonID:} See {\it LogonID}.
\item {\it Status:} Status of the authentication request. ``0x0'' means success otherwise failure, see \cite{ultimatewindows} for failure codes for the appropriate Event ID.
\item {\it Source:} For authentication events, this will correspond to the the computer where the authentication originated (source computer), if it is a local logon event then this will be the same as the {\it LogHost}.
\item {\it ServiceName:} The account name of the computer or service the user is requesting the ticket for.
\item {\it Destination:} This is the server the mapped credential is accessing. This may indicate the local computer when starting another process with new account credentials on a local computer.
\item {\it AuthenticationPackage:} The type of authentication occurring including Negotiate, Kerberos, NTLM plus a few more.
\item {\it FailureReason:} The reason for a failed logon.
\item {\it ProcessName:} The process executable name, for authentication events this is the process that processed the authentication event. ProcessNames may include the file type extensions (i.e exe).
\item {\it ProcessID:}  A semi-unique (unique between currently running processes AND {\it LogHost}) value that identifies the process.  Process ID allows you to correlate other events logged in association with the same process through to the process end.
\item {\it ParentProcessName:} The process executable that started the new process. ParentProcessNames often do not have file extensions like ProcessName but can be compared by removing file extensions from the name.
\item {\it ParentProcessID:}  Identifies the exact process that started the new process. Look for a preceding event 4688 with a {\it ProcessID} that matches this {\it ParentProcessID}.
    \end{itemize}

\section{Event Attributes}\label{app:attributes}

  \begin{table}[h!]{
    
  \centering
  \begin{tabular*}{1\textwidth}{@{\extracolsep{\fill}} p{0.65\textwidth} l}
    \toprule[\heavyrulewidth]
    Event Ids & Attribute\\
    \midrule[\heavyrulewidth]
    All & Time\\
    All & EventID\\
    All & LogHost\\
    4624, 4625, 4634 & LogonType \\
    4624, 4625, 4634 & LogonTypeDescription \\
    All except System Events & UserName\\
    All except System Events & DomainName\\
    All except 4768, 4769, 4770, 4774, 4776 & LogonID\\
    4624 (LogonType 9), 4648, 4774 & SubjectUserName\\
    4624 (LogonType 9), 4648, 4774 & SubjectDomainName\\
    4624 (LogonType 9), 4648 & SubjectLogonID\\
    4768, 4769, 4776 & Status\\
    4624, 4625, 4648, 4768, 4769, 4770, 4776 & Source\\
    4769, 4770 & ServiceName\\
    4648 & Destination\\
    4624, 4625, 4776 & AuthenticationPackage\\
    4625 & FailureReason\\
    4624, 4625, 4648, 4688, 4689 & ProcessName\\
    4624, 4625, 4648, 4688, 4689 & ProcessID\\
    4688 & ParentProcessName\\
    4688 & ParentProcessID\\
    \bottomrule[\heavyrulewidth]
    \end{tabular*}}
    \caption{Event Attributes}
  \label{tab:field_desc}
\end{table}

\bibliographystyle{plainnat}
\bibliography{data_arxiv}

\end{document}